# Spatio-Temporal Dynamics of Visual Imagery for Intuitive Brain-Computer Interface

Seo-Hyun Lee[1], Minji Lee[1], Seong-Whan Lee[1, 2]
[1]Department of Brain and Cognitive Engineering, Korea University, Seoul, Republic of Korea
[2]Department of Artificial Intelligence, Korea University, Seoul, Republic of Korea
seohyunlee@korea.ac.kr, minjilee@korea.ac.kr, sw.lee@korea.ac.kr

*Abstract*—Visual imagery is an intuitive brain-computer interface paradigm, referring to the emergence of the visual scene. Despite its convenience, analysis of its intrinsic characteristics is limited. In this study, we demonstrate the effect of time interval and channel selection that affects the decoding performance of the multi-class visual imagery. We divided the epoch into time intervals of 0-1 s and 1-2 s and performed six-class classification in three different brain regions: whole brain, visual cortex, and prefrontal cortex. In the time interval, 0-1 s group showed 24.2 % of average classification accuracy, which was significantly higher than the 1-2 s group in the prefrontal cortex. In the three different regions, the classification accuracy of the prefrontal cortex showed significantly higher performance than the visual cortex in 0-1 s interval group, implying the cognitive arousal during the visual imagery. This finding would provide crucial information in improving the decoding performance.

## I. Introduction

Brain-computer interface technology is developing in a more convenient and intuitive way to benefit not only patients but also healthy people [1-8]. Recently, there have been attempts towards investigating an intuitive brain-computer interface (BCI) paradigm that can operate only by the user's intrinsic thought [9, 10]. Visual imagery is one of the recent paradigms that hold potential towards an intuitive BCI control [11]. It refers to the emergence of a constructive representation of the visual scene, without exposure to any external stimuli [12]. Unlike the conventional BCI paradigms such as steady-state visual evoked potential or event-related potential [13-16], visual imagery can be operated without any external stimuli. It is an endogenous paradigm that provides freedom and convenience to the user [17]. When using the visual imagery paradigm, there is no need for another procedure than imagining the visual scene of the object. For example, if you imagine an apple, then the system is able to decode the apple in one step [11].

Although visual imagery holds intuitiveness and effectiveness, it still is a new emerging paradigm reporting relatively low decoding performance. Kosmyna et al. [12] showed the presence of visual imagery by acquiring 77 % in discriminating visual imagery state versus resting state. However, they reported 55.9 % of binary classification accuracy, failing to discriminate between two different classes of visual imagery. Lee et al. [11] showed the potential of multiclass classification of visual imagery by reporting 22.2 % of thirteen-class classification accuracy, however, the state-of-the-art decoding performance is still remaining at a low level. This can be due to the unknown features and decoding methods that fit well into the visual imagery paradigm. In contrast, another endogenous paradigm, motor imagery (MI), is actively studied and recording relatively robust classification performance. In the case of MI paradigm, the key components such as event-related desynchronization or event-related synchronization are predominated [18]. Also, the specific frequency band of 8-14 Hz and the relevant brain region is also revealed, leading to empirical research on the MI paradigm [19, 20]. When referring to this, visual imagery also holds the potential to be robustly decoded when taking its peculiar features into account.

Visual imagery is known to be processed in the occipital region of the brain [21], however, some research reported the changes in the alpha power in the prefrontal cortex while performing the visual imagery [17]. Ishai et al. [22] discovered that visual imagery originates from top-down mechanisms arising from prefrontal cortex. Dijkstra et al. [21] reported that the representations of visual imagery arises in different temporal dynamics compared to perception. Yet, brain state alteration during the visual imagery in different time epoch and brain region haven't been reported. Although brain modeling and functional magnetic resonance imaging studies about visual imagery are actively proposed, the direct impact of the neural background affecting the decoding performance in a practical view of BCI is yet discovered [23, 24]. In this paper, we investigate the effect of time intervals and channel selection on the visual imagery decoding using electroencephalography (EEG). We observed the brain state changes in the first and the last seconds of visual imagery and analyzed each classification performance in different brain regions. The decoding performance of each group was compared in respect of the visual imagery processing procedure. Our research could show the temporal dynamics of visual imagery, therefore, contribute to the improvement of the decoding performance.

This work was supported by Institute for Information & Communications Technology Promotion (IITP) grant funded by the Korea government (No. 2017-0-00451, Development of BCI based Brain and Cognitive Computing Technology for Recognizing User's Intentions using Deep Learning).

## II. METHOD

### A. Data Acquisition

Nine healthy subjects were recruited for the experiment (age: 25.56 ± 2.35, all males). Every subject had no history of neurological disease, nor visual impairment. All subjects signed the informed consent according to the Helsinki declaration. The experimental protocols and environments were reviewed and approved by the Institutional Review Board at Korea University [KUIRB-2019-0143-01].

Subjects were seated on a chair with a visual display monitor, providing a cue for performing the visual imagery. The subjects were instructed to imagine the given visual cue four times in a row, for 2 s period. The given cue was images representing twelve different words ('ambulance', 'clock', 'hello', 'help me', 'light', 'pain', 'stop', 'thank you', 'toilet', 'TV', 'water', and 'yes'). The chosen words are generally used for the patients' basic communication [25]. Before each cue, 3 s of resting time was provided. The visual cue of twelve words or a rest class were given for the following 2 s. After the cue and before each visual imagery onset, 0.8-1.2 s of cross mark appeared on the screen. The subjects were instructed to imagine the given visual scene as soon as the cross mark disappeared on the screen. 100 trials of visual imagery data were recorded per every twelve words and rest class.

The EEG data were collected via Brain Vision/Recorder (BrainProduct GmbH, Germany) using 64-channel EEG cap with active Ag/AgCl electrodes. The electrode placements followed the international 10-20 system. FCz channel was used as a reference electrode, and FPz channel was used for the ground electrode. The impedances of the electrodes were maintained below 10 Hz during the experiment.

### B. Data Analysis

The data was down-sampled to 256 Hz and band-pass filtered to the frequency range of 1-100 Hz using the 5th order Butterworth filter. The visual imagery signals were segmented to the imagery performed 2 s epochs, and baseline corrected by subtracting the average value of -200 ms of the imagery onset cue. Among the twelve words and rest class, we analyzed six-word classes that have specific shapes. We selected the epochs of six words ('ambulance', 'clock', 'light', 'toilet', 'TV', and 'water') from the whole data. Each epoch was again segmented to 0-1 s and 1-2 s interval groups.

We performed six-class classification of visual imagery in the two time interval groups (0-1 s, 1-2 s) within three channel selection groups (64-channel group, visual cortex group, prefrontal cortex group). The classification groups were designed in order to analyze the alteration of brain state in different brain areas during the imagination. The two time interval groups were divided into first and last 1 s of visual imagery phase without any overlaps to identify the significantly different features of the onset and the ongoing state of the imagery. The visual cortex group consisted of nine channels in the occipital lobe. The prefrontal cortex group contained nine channels in the frontal lobe. We used the first and last three common spatial pattern (CSP) features for the classification in the 64-channel group. Only the first and last CSP features were used for the nine-channel classification group, in order to consider the number of features for the projection. We used one-versus-rest (OVR) strategy for the multiclass CSP. OVR strategy separated each class from all the rest classes and performed binary CSP since CSP is originally a binary feature extraction method [26].

The classification was performed by the shrinkage regularized linear discriminant analysis (RLDA). The shrinkage RLDA classifier adds a regularization term to a covariance matrix using optimal shrinkage parameter [27]. We performed 10-fold cross-validation by randomly separating the test set from the training set in 1:9 ratio [28]. The training and test were performed 10 times. The final classification accuracy was calculated as the average value of 10 times of classification results. The classification accuracy of the two time interval groups and three channel selection groups were compared with statistical analysis. For the comparison of the classification accuracy between two time interval groups, as well as two channel selection groups, we performed non-parametric Bootstrap analysis as a statistical analysis method [29]. The significance level was set at $\alpha = 0.05$.

## III. RESULTS

### A. Comparison of the Two Time Interval Groups

Table I shows the classification result of 64-channel group in the two time intervals. Table II and Table III presents the classification performance of the visual cortex group and the prefrontal cortex group. The average 10-fold cross-validation accuracy of 64-channel group was 25.9 % and 25.3 % in the 0-1 s and 1-2 s interval group, respectively. In the visual cortex group, the average accuracy of nine subjects was 21.0 % and 20.2 % in the 0-1 s group and 1-2 s group, respectively. The average classification performance of the prefrontal cortex group was 24.2 % in 0-1 s group and 22.5 % in 1-2 s interval group. Two time interval groups showed no significant difference between each other in all channel group ($t = 0.62$, $p$-value = 0.45), and visual cortex group ($t = 1.33$, $p$-value = 0.085). However, the average 10-fold cross-validation accuracy of 0-1 s interval group showed significantly higher performance compared to the 1-2 s interval group in the prefrontal cortex ($t = 3.07$, $p$-value $< 0.001$).

TABLE I. THE CLASSIFICATION ACCURACY (%) OF ALL 64-CHANNEL GROUP IN 0-1 S AND 1-2 S TIME INTERVALS

|  | Time interval | |
|---|---|---|
|  | *0 – 1 s* | *1 – 2 s* |
| Subject 1 | 25.3 ± 1.3 | 25.0 ± 1.0 |
| Subject 2 | 23.3 ± 1.7 | 25.8 ± 1.1 |
| Subject 3 | 26.7 ± 0.9 | 25.0 ± 1.4 |
| Subject 4 | 24.2 ± 1.2 | 23.1 ± 0.7 |
| Subject 5 | 25.1 ± 1.4 | 26.4 ± 1.1 |
| Subject 6 | 23.0 ± 1.0 | 24.3 ± 1.1 |
| Subject 7 | 21.1 ± 1.5 | 20.3 ± 1.6 |
| Subject 8 | 37.1 ± 1.2 | 32.1 ± 1.3 |
| Subject 9 | 27.1 ± 1.0 | 26.5 ± 0.9 |
| **Average** | **25.9 ± 4.6** | **25.3 ± 2.7** |

TABLE II.   THE CLASSIFICATION ACCURACY (%) OF VISUAL CORTEX GROUP IN 0-1 S AND 1-2 S TIME INTERVALS

|  | Time interval | |
| --- | --- | --- |
|  | *0 – 1 s* | *1 – 2 s* |
| Subject 1 | 23.0 ± 1.4 | 21.9 ± 1.0 |
| Subject 2 | 22.4 ± 1.1 | 20.0 ± 1.1 |
| Subject 3 | 19.3 ± 1.5 | 21.5 ± 1.4 |
| Subject 4 | 18.2 ± 1.3 | 18.9 ± 0.7 |
| Subject 5 | 19.5 ± 1.0 | 18.8 ± 1.1 |
| Subject 6 | 22.3 ± 1.8 | 20.7 ± 1.1 |
| Subject 7 | 21.3 ± 0.9 | 17.5 ± 1.6 |
| Subject 8 | 22.5 ± 0.7 | 20.8 ± 1.3 |
| Subject 9 | 20.5 ± 1.5 | 21.5 ± 0.9 |
| **Average** | **21.0 ± 1.7** | **20.2 ± 1.5** |

TABLE III.   THE CLASSIFICATION ACCURACY (%) OF PREFRONTAL CORTEX GROUP IN 0-1 S AND 1-2 S TIME IINTERVALS

|  | Time interval | |
| --- | --- | --- |
|  | *0 – 1 s* | *1 – 2 s* |
| Subject 1 | 23.4 ± 1.3 | 19.6 ± 1.1 |
| Subject 2 | 23.2 ± 0.9 | 23.8 ± 1.1 |
| Subject 3 | 25.1 ± 1.5 | 22.2 ± 0.6 |
| Subject 4 | 24.7 ± 1.1 | 22.6 ± 1.1 |
| Subject 5 | 22.6 ± 0.7 | 23.7 ± 1.3 |
| Subject 6 | 23.0 ± 0.8 | 19.8 ± 0.8 |
| Subject 7 | 19.7 ± 1.0 | 18.5 ± 0.7 |
| Subject 8 | 34.7 ± 1.2 | 32.6 ± 0.5 |
| Subject 9 | 21.4 ± 1.2 | 19.8 ± 0.9 |
| **Average** | **24.2 ± 4.3** | **22.5 ± 4.2** |

Figure 1 shows the confusion matrix of the six-class classification on 0-1 s and 1-2 s interval groups for subject 8. Subject 8 showed the highest classification performance in all 64-channel group, therefore, was chosen as a representative. The confusion matrix for 0-1 s time interval shows a relatively large difference in the true-positive rate among classes. While 'ambulance' had 31.2 % of true positive rate, 'water' showed 13.1 % lower true positive rate of 18.1 %. However, in the confusion matrix of 1-2 s interval group, the gap in the true positive rate among classes decreased to 8.4 %. The 'light' showed the highest true positive rate of 29.2 %, while the lowest 'toilet' had 20.8 %. The standard deviation of the six-class true positive rate was 5.20 % in 0-1 s interval group, and 3.10 % in the 1-2 s interval group. Figure 2 shows the CSP pattern of the six-class visual imagery in 0-1 s and 1-2 s interval groups. While the distinct brain region shows biased to the 'ambulance' in 0-1 s interval group, a distinct brain region shown in 1-2 s interval group is relatively equally distributed upon classes.

### B.  Comparison of the Channel Selection Performance

Figure 3 and Figure 4 shows the comparison of the classification accuracy of the visual cortex and prefrontal cortex in 0-1 s time interval. The average classification accuracy of prefrontal cortex was 21.0 ± 1.7 % which was significantly higher than the visual cortex, 24.2 ± 4.3 % ($t = -2.24$, $p$-value $< 0.001$). However, the classification accuracy of the visual cortex and prefrontal cortex in 1-2 s time interval was 20.2 ± 1.5, and 22.5 ± 4.2 % respectively. The two channel selection groups in the 1-2 s time interval did not show significant difference between each other ($t = -1.60$, $p$-value = 0.05).

### IV.   DISCUSSION AND CONCLUSION

Our result showed significantly higher classification accuracy in 0-1 s interval group than the 1-2 s group in the prefrontal cortex. Also, the classification performance in the prefrontal cortex showed significantly higher accuracy than the visual cortex in 0-1 s interval group. The standard deviation of the six-class true positive rate in subject 8 was higher in 0-1 s interval group when using the whole channel. The CSP pattern of subject 8 showed biased aspect in 0-1 s interval group compared to the 1-2 s interval group, resulting in the larger gap among the true positive rate.

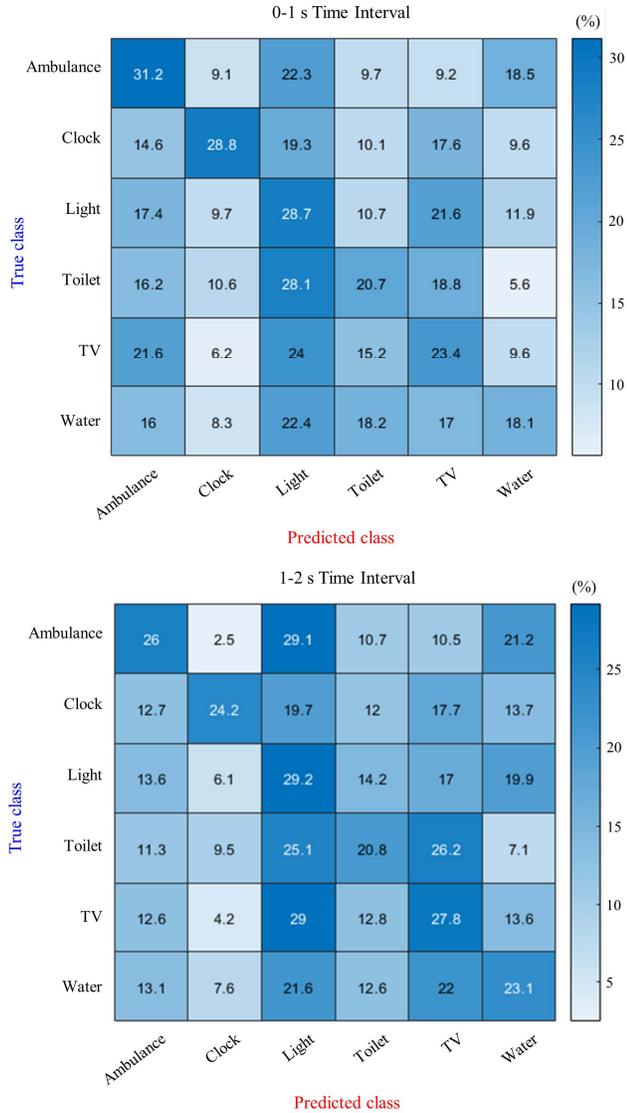

Figure 1. The confusion matrix of 0-1 s time interval (top) and 1-2 s time interval (bottom) in all 64 channel group for subject 8. The classification was performed by CSP feature and RLDA shrink classifier.

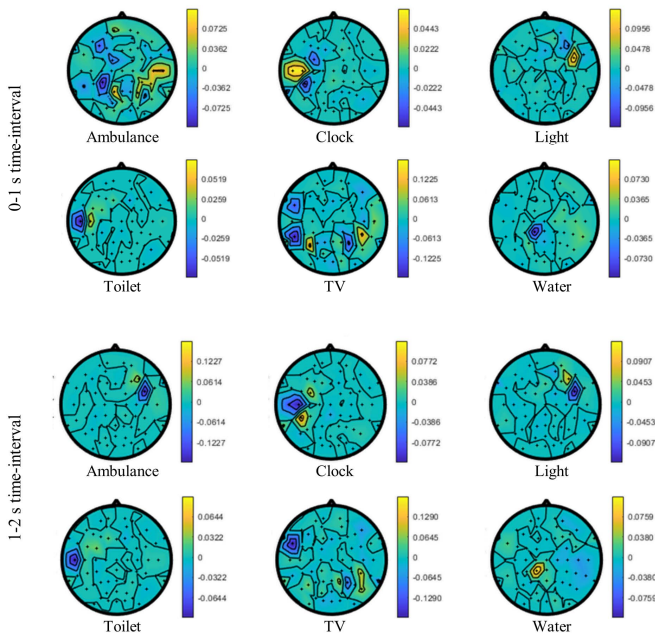

Figure 2. The first and the last CSP patterns of 0-1 s interval group (top) and 1-2 s interval group (bottom) for subject 8 in discriminating six-class visual imagery.

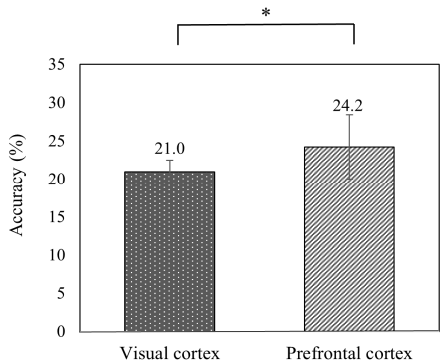

Figure 3. Comparison of the classification accuracy of visual cortex group and prefrontal cortex group in 0-1 s interval group (* < 0.05). Classification was performed by CSP feature and RLDA shrink classifier.

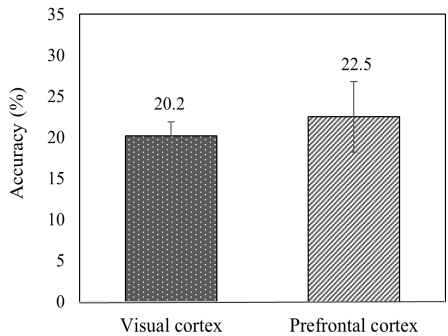

Figure 4. Comparison of the classification accuracy of visual cortex group and prefrontal cortex group in 1-2 s interval group. Classification was performed by CSP feature and RLDA shrink classifier. Error bar indicates the standard deviation of the classification accuracy.

The visual imagery process occurs initially in the prefrontal cortex, and then move to another brain region [22]. Our findings support this mechanism, the visual imagery arising from the prefrontal cortex. The prefrontal cortex is known to play an important role in cognitive control and holds the ability to orchestrate thought and action in accordance with internal goals [30]. Therefore, we could infer that the prefrontal cortex is highly activated during the first few seconds of the visual imagery, reflecting the user intention and the semantic meaning of the emerging object [31]. The result can also be supported by Kosyma et al. [12] reporting the significantly higher alpha activity in 770 ms to 830 ms. The classification accuracy of two time interval groups of visual cortex did not show significant differences, implying that the visual cortex is involved in the whole processing period [32].

The true-positive rate of different classes differed in a larger gap in the 0-1 s interval group than the 1-2 s group. Therefore, we could infer that the visual scene of different words may have different arousal time points in the brain. Also, the CSP pattern plot shows that the class with a higher true-positive rate shows distinctive patterns compared to the classes with the low true positive rate. According to our result, it can be interpreted that the class showing distinct brain area is relatively well discriminated against than the other classes. Interestingly, the CSP plot also showed distinct patterns in the parietal region in some classes. This phenomenon is in line with Dentico et al. [23] reporting an increased top-down signal flow in parieto-occipital cortices during visual imagery.

Further analysis of the spatial representations of visual imagery by subdividing the brain regions may contribute to a thorough comprehension of the paradigm. In addition, alpha power increase in the prefrontal cortex is reported in Sousa et al. [17]. Also, Kosmyna et al. [12] reported significantly high alpha activity in the visual cortex. Further research focusing on the alpha power changes according to different time intervals may contribute to appropriate feature selection of visual imagery, therefore, enhancing the decoding performance [33]. Also, further analysis using methods such as wavelet transform and more segmentalized time interval considering the local information may contribute to exhausted understanding the visual imagery paradigm [34, 35]. Overall, our research could suggest an insight into more precise comprehension of the visual imagery paradigm, leading to practical use of visual imagery in the real world.